\documentclass[
 preprint,
]{aastex62}

\usepackage{graphicx}
\usepackage{dcolumn}
\usepackage{bm}
\usepackage{color}
\usepackage{amsmath, amssymb}
\received{January 1, 2019}
\revised{January 7, 2019}
\accepted{\today}
\submitjournal{ApJ}

\shorttitle{Evolution of three-dimensional Relativistic Ion Weibel Instability}
\shortauthors{Takamoto et al.}

\begin{document}

\title{Evolution of three-dimensional Relativistic Ion Weibel Instability: Competition with Kink Instability}

\correspondingauthor{Makoto Takamoto}
\email{mtakamoto@eps.s.u-tokyo.ac.jp}

\author{Makoto Takamoto}
 \affiliation{Central Research Laboratories, NEC Corporation, 1753 Shimonumabe, Nakahara-ku, Kawasaki city, Kanagawa, 211-8666, Japan}
\affil{Department of Earth and Planetary Science, The University of Tokyo, Hongo, Bunkyo-ku, Tokyo 113-0033, Japan}

\author{Yosuke Matsumoto}
\affiliation{Department of Physics, Chiba University 1-33 Yayoi-cho, Inage-ku, Chiba 263-8522, Japan}

\author{Tsunehiko N. Kato}
\affiliation{Center for Computational Astrophysics, National Astronomical Observatory of Japan, 2-21-1 Osawa, Mitaka, Tokyo 181-8588, Japan}

\begin{abstract}
In this paper, we report our recent findings on the relativistic Weibel instability
and its nonlinear saturation by performing numerical simulations of collisionless plasmas. 
Analysis of the obtained numerical results revealed that 
the nonlinear phase of the Weibel instability can be described 
by characteristic phases based on the Weibel filaments' current density in terms of particle and Alfv\'en limit currents. 
We also analyzed the relativistic kink instability based on the energy principle in the magnetohydrodynamic (MHD) regime, 
and found that 
the Weibel filaments do not suffer from the kink-type instability in the MHD regime up to 1000 $\omega_{p,i}^{-1}$. 
This finding allowed a magnetic field to be sustained by relativistic Weibel instability 
that was stable enough to be a seed for MHD dynamos. 
\end{abstract}

\keywords{editorials, notices --- 
miscellaneous --- catalogs --- surveys}


\section{\label{sec:sec1}Introduction}

Weibel instability converts the anisotropy of plasma temperature into a magnetic field \citep{1959PhRvL...2...83W,1959PhFl....2..337F}, 
and is considered as an origin of magnetic field in various phenomena. 
In particular, 
it is expected that 
magnetic field turbulence resulting from relativistic Weibel instability is responsible for the 
relativistic collisionless shock formation 
\citep{2008ApJ...681L..93K,2008ApJ...673L..39S}, 
which is one of the most probable candidates for accelerating the ultra-high-energy cosmic-rays \citep{1984ARA&A..22..425H}
through diffusive shock acceleration \citep{1999JPhG...25R.163K,2001MNRAS.328..393A} and wake field acceleration \citep{1979PhRvL..43..267T,2008ApJ...672..940H,2017ApJ...840...52I,2018ApJ...858...93I}. 
And Weibel instability is also regarded as providing the necessary magnetic field to explain the observed afterglow of the gamma-ray burst \citep{2006ApJ...642..389N}. 

For the reasons mentioned above, 
the relativistic Weibel instability has been studied extensively under various situations and parameters \citep{2005PhPl...12h0705K,2006PhRvL..96j5008C,2007AA...475....1A,2007AA...475...19A,2007PPCF...49.1885P}. 
Importantly, 
recent developments in high-power laser facilities allow direct investigations of the Weibel instability in laboratory.
In particular, 
the Omega and NIF laser experiments have recently performed several projects 
which are considered to have observed the early phase development of nonrelativistic Weibel filaments~\citep{2015NatPh..11..173H,2015PhPl...22e6311P,2017PhPl...24d1410H}. 
Although these recent laser experiments have not yet been successful in realizing collisionless shock formation, 
we expect that 
the next-generation laser facilities would succeed in achieving fully developed relativistic collisionless shock via the relativistic Weibel instability. 

In this paper, 
we report the further analysis of our previous work on the relativistic Weibel instability \citep{2018ApJ...860L...1T} (TMK18 in the following). 
The Weibel instability is a very popular mechanism in plasma instability and has been studied extensively, 
including analytical and numerical approaches. 
However, these work were limited to either only the early phase or to the two-dimensional (2D) system. 
In our work, 
large-scale and long-time simulations of the relativistic Weibel instability were performed. 
In Section 2, 
the numerical setup is explained. 
In Section 3, numerical results and our analyses are presented. 
In Section 4, 
the kink instability in the Weibel filaments is discussed. 
Section 5 discusses the saturation mechanism of the Weibel instability, 
and Section 6 summarizes our results. 

\section{\label{sec:sec2}Numerical Setups}

In this paper, 
the temporal evolution of the relativistic ion-electron Weibel instability is modeled by a collisionless plasma; 
it is integrated by a three-dimensional (3D) relativistic electromagnetic particle-in-cell (PIC) code developed by \citet{matsumoto2017}, 
which is a vectorized, hybrid-parallel 3D PIC simulation code with a quadratic particle weighting function. 
As considered in TMK18, 
two cold, $T = 0$, unmagnetized beam counter-flows are modeled in our 2D and 3D simulation boxes. 
The beam is set in the x-direction; 
in each cell, 
half of the particles are set to move in the positive x-direction, 
and the other particles move in the negative x-direction. 
The periodic boundary is considered in all directions. 
Consequently, 
a relativistic ion-electron Weibel instability grows in the whole numerical box region because of the very large anisotropy. 
For a unit of length and time, 
the nonrelativistic ion inertial length $c/\omega_{p,i}$ and the inverse of the ion plasma frequency $\omega_{p,i}$ are considered. 
To resolve the initial electron Weibel instability accurately, 
the cell size is set as $\Delta = 0.1 c/\omega_{p,e}$, equivalent to $0.07 c/\tilde{\omega}_{p,e}$ in the case of $\gamma = 2$, 
$0.045 c/\tilde{\omega}_{p,e}$ in the case of $\gamma = 5$, 
and $0.022 c/\tilde{\omega}_{p,e}$ in the case of $\gamma = 20$, 
where $\tilde{\omega}_{p,e}$ is the relativistic electron plasma frequency. 
To reduce the numerical Cherenkov instability, the CFL number is set to unity, $\Delta t = \Delta/c$, 
which is a magical time-step size specific to our semi-implicit PIC code \citep{2015PASJ...67...64I}. 
Simulation parameters of the initial beam bulk Lorentz factor $\gamma$, the ion-to-electron mass ratio $M/m$, 
number of particles per cell per species $n_0$, 
and the system size ($L_{\rm x}$, $L_{\rm y}= L_{\rm z} = L_{\perp}$) are summarized in table \ref{table1} for different simulation runs. 

\begin{table}
 \centering
  \caption{List of the parameters. $\gamma$ is the initial Lorentz factor of the particles, and $M, m$ are ion and electron masses, respectively; 
           $n_0$ is the initial particle number per cell; 
           $L_{\rm x}, L_{\perp}$ are the numerical box size in parallel and perpendicular to the initial beam direction. }
  \begin{tabular}{lcccccl}
  \hline
    Name & \ $\gamma$ \  & \ $M/m$ \ & \ $n_0$ \ & \ $L_{\rm x} \omega_{p,i}/c$ \ & $L_{\perp} \omega_{p,i}/c$ \  \\
  \hline
    runA     &  2 &  100 & 20 & 18 & 15.36  \\
    runAa    &  2 &  100 & 20 & 18 &  6     \\
    runAb    &  2 &  100 & 20 & 18 &  1     \\
    runB     &  5 &  100 & 20 & 18 & 15.36  \\
    runB1    &  5 &    4 & 20 & 18 & 15.36  \\
    runB2    &  5 &   25 & 20 & 18 & 15.36  \\
    runB3 (2D,x-z) &  5 & 1836 & 20 & 49 & 95.6   \\
    runBa (TMK18)    &  5 &   25 & 10 & 20 & 51.2   \\
    runC     & 20 &  100 & 20 & 18 & 15.36  \\
  \hline
\end{tabular}
\label{table1}
\end{table}

\section{\label{sec:sec3}Results}

\subsection{\label{sec:sec3.1}Theoretical Consideration of Current Evolution}

In this section, 
we briefly review the current evolution of Weibel filaments 
reported by \citet{2005PhPl...12h0705K} in the case of a 2D (out-of-plane current) study. 

In the early nonlinear phase, 
the Weibel instability is known to generate current filaments because of the generated magnetic field, 
that is, two lines of current flowing in the same direction attract each other, 
but those in the opposite direction repel each other, 
resulting in an accumulation of currents. 

In the later nonlinear phase, 
\citet{2005PhPl...12h0705K} pointed out that 
there are two possible upper limits of current; 
One is the \textit{Alfv\'en limit current}~\citep{1939PhRv...55..425A}, 
which results in the magnetic field for which the gyro-radius of the particles is equivalent to the current filaments' radius. 
In Gaussian units, 
this is given as: 
\begin{equation}
  \label{eq:3.1.1}
  I_{\rm A} = I_0 \langle \gamma \beta_{||} \rangle
  ,
\end{equation}
where $I_0 \equiv m_{s} c^3/q_{s}$, 
$m_{s}$ and $q_{s}$ are the mass and charge of the particles, respectively, 
and $s$ means ion or electrons. 
$\beta_{||}$ is the magnitude of the three-velocity in units of light velocity along the filaments' direction, 
and $\gamma$ is the Lorentz factor. 
The angle brackets $\langle \rangle$ denote the average over the beam volume. 
This gives a theoretically maximal magnetic field~\citep{1939PhRv...55..425A,1999ApJ...526..697M}.
The other limit is the \textit{particle limit current}, 
which is the maximum current $I_{\rm P}$ carried by all the particles in the beam, 
defined as: 
\begin{equation}
  \label{eq:3.1.2}
  I_{\rm P} = \pi R^2 J_{\rm P}, \quad J_{\rm P} \equiv e \mu n_s \langle \beta_{||} \rangle c
  ,
\end{equation}
where $R$ is the radius of filaments, 
$n_s$ is the density of particles, 
and $\mu = 1$ for electron-proton plasmas, and $\mu = 2$ for electron-positron plasma. 

Following these upper current limits, 
\citet{2005PhPl...12h0705K} pointed out that 
there are two paths of evolution depending on the initial anisotropy of particle momentum:  
i.e., the \textit{Alfv\'en limit} ($I_{\rm A} < I_{\rm P}$), and the \textit{particle limit} ($I_{\rm A} > I_{\rm P}$). 
The Alfv\'en limit occurs 
when the initial anisotropy is small or the beams have a nonrelativistic velocity. 
In this case, 
the currents by Weibel instability saturates at the Alfv\'en current, 
and increase its radius without changing its current, 
resulting in a gradually decreasing magnetic field. 
In contrast, 
the particle limit occurs 
when the initial anisotropy is large or for beams with relativistic velocity. 
In this case, 
currents continue to evolve by coalescing 
until reaching the Alfv\'en current. 

\citet{2005PhPl...12h0705K} studied these saturation mechanisms using a 2D relativistic PIC simulation, 
and reported a detailed analysis. 
However, 
it is still unclear whether these results can be applied in the 3D case 
because there are several important instabilities that occur only when a 3D case is considered, 
such as the kink instability. 
In the following, 
we investigate the saturation mechanism of the Weibel instability in a 3D space. 

\subsection{\label{sec:sec3.2}Temporal Evolution of Magnetic Field and Current}

\begin{figure}[t]
 \centering
  \includegraphics[width=8.cm,clip]{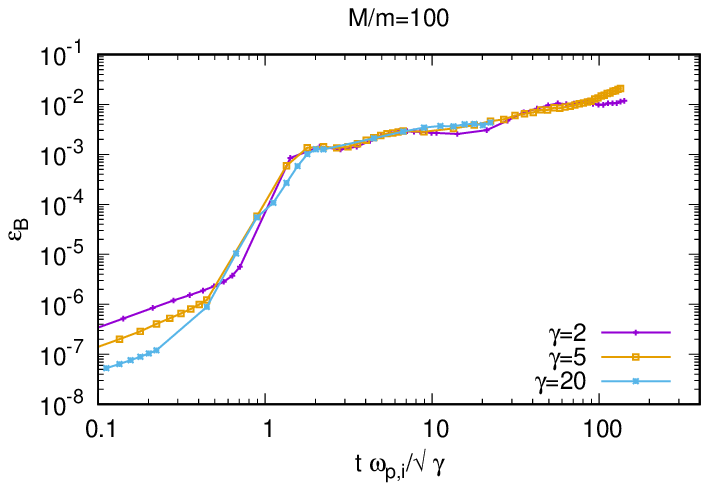}
  \includegraphics[width=8.cm,clip]{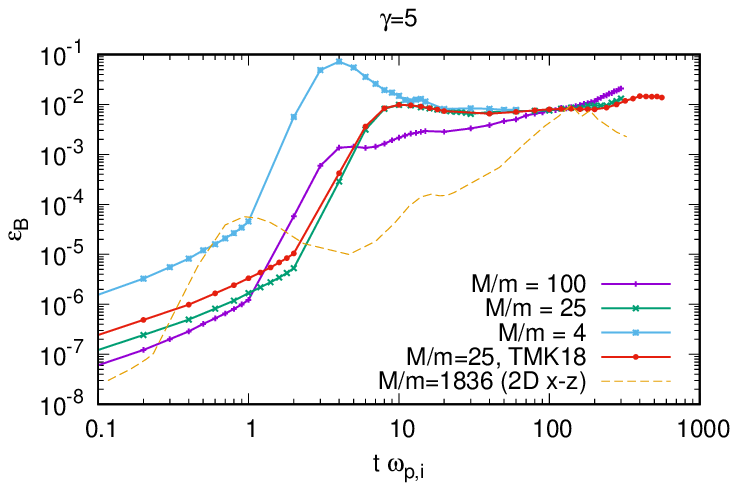}
  \caption{Temporal evolution of $\epsilon_B \equiv B^2/8\pi M n c^2 (\gamma - 1)$. 
           Left: Lorentz factor dependence in terms of time scale in the particle comoving frame: $t/\sqrt{\gamma}$; 
           Right: Mass ratio dependence in terms of time scale in the laboratory frame. 
           }
  \label{fig:3.2.1}
\end{figure}

Figure \ref{fig:3.2.1} shows the temporal evolution of the magnetic field generated by Weibel instability, 
which is measured by 
\begin{equation}
  \label{eq:3.2}
  \epsilon_{\rm B} \equiv \frac{B^2}{8 \pi M n c^2 (\gamma - 1)}
  ,
\end{equation}
where $M$ is the ion mass, $n$ is the initial ion number density in laboratory frame, $c$ is the light velocity, 
and $\gamma$ is the Lorentz factor of the initial ion. 
The term $\epsilon_{\rm B}$ measures the energy density of the magnetic field 
with respect to the initial ion kinetic energy density, 
and the plotted values are averaged over the entire volume. 
The left panel shows the Lorentz factor dependence of the generated magnetic field in terms of time scale in the  particle comoving frame, 
plotting the runs A, B, C 
($\gamma = 2, 5, 20$).
It shows that 
the generated magnetic is insensitive to the plasma Lorentz factor. 
This indicates that 
the generated magnetic field can be described by Amp\`ere's law: $B \sim 4\pi n q (c/\tilde{\omega}_{\rm p,i})$ 
which makes the magnetization parameter $\epsilon_{\rm B}$ a constant value
\footnote{
We consider that 
the differences in the linear phase reflect those in the spatial resolution because of the Lorentz contraction. 
}. 

The right panel shows the mass dependence of the magnetic field evolution. 
Although they show a very different evolution before $10 \omega_{\rm p,i}^{-1}$, 
the saturation values after $10 \omega_{\rm p,i}^{-1}$ take a similar value around 0.01 to 0.02. 
This can be understood from the current evolution discussed in the following. 

\begin{figure}[t]
 \centering
  \includegraphics[width=8.cm,clip]{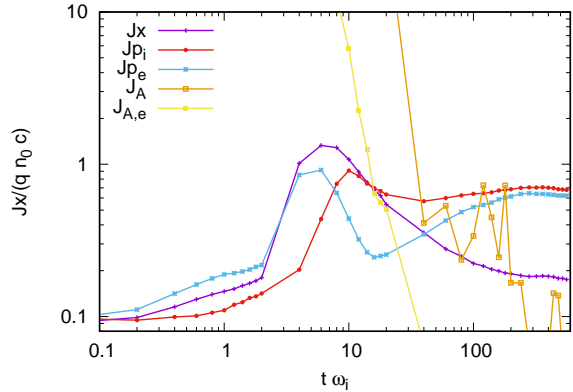}
  \caption{Temporal evolution of current density in the x-direction in the case of runBa. 
           $J_{\rm x}$ is the current density measured at $x = L_{\rm x}/2$ averaged over this plane, 
           $J_{\rm p,i}$ is the ion particle limit current, 
           $J_{\rm p,e}$ is the electron particle limit current, 
           $J_{\rm A}$ is the electron Alfv\'en limit current, 
           and $J_{\rm A,e}$ is the electron Alfv\'en limit current, 
           }
  \label{fig:3.2.2}
\end{figure}

Figure \ref{fig:3.2.2} is the temporal evolution of current density in the x-direction in the case of runBa, 
where the plotted values are measured at $x = L_{\rm x}/2$ averaged over this plane
\footnote{
In Figure \ref{fig:3.2.2}, 
both the particle and the Alfv\'en limit currents were calculated using Equations (\ref{eq:3.1.1}) and (\ref{eq:3.1.2}). 
The physical variables were measured at a plane on $x=L_{\rm x}/2$ and were averaged over that plane. 
Concerning the $\gamma$ factor in the Alfv\'en limit current, 
the initial Lorentz factor was used. 
}
. 

This indicates that 
the temporal evolution can be divided into six phases, i.e.: 
(1) a linear electron Weibel growth phase, 
(2) nonlinear phase of electron Weibel instability following the electron particle limit current, 
(3) a linear ion Weibel growth phase, 
(4) a nonlinear ion Weibel growth phase following the ion particle limit current, 
(5) a coalescing phase of Weibel filaments, 
(6) a saturation phase due to either the Alfv\'en limit current (runBa) or limited numerical box size. 
Phase (3) can be observed more clearly in run B3, 
where $\epsilon_{\rm B}$ initially increases exponentially 
because of the growth in the electron Weibel instability (1) and (2)
, and changes its growth rate around $t = 0.3 \omega_{p,i}^{-1}$. 
In the other runs, 
although the phases (1) and (2) exist, 
the smaller value of mass ratio causes these regions to degenerate into the linear growth phase of ion Weibel instability. 
Around $t = 4 \omega_{p,i}^{-1}$, 
the current density becomes larger than the electron particle limit, 
and instead starts to follow the ion particle limit around $t = 10 \omega_{p,i}^{-1}$, 
which coincides with the first peak of $\epsilon_{\rm B}$ other than runB3 
whose large mass ratio allows the electron particle limit to show a clear peak. 
After that, 
the coalescence of the filaments starts, 
and the filaments increase their radius as discussed later. 
At this phase, 
the current density does not follow the ion particle limit current density 
and its value starts to decrease 
because the current reaches the electron Alfv\'en limit 
and electrons can no longer participate in increasing current. 
Finally, at $t \gtrsim 300\omega_{p,i}^{-1}$ of runBa, 
the current density also reaches the ion Alfv\'en current, 
which also limits the ion contribution to the current, 
resulting in a reduced current density. 
The $\epsilon_{\rm B}$ reaches the saturation value at around $0.02$. 
In our 3D runs, 
we did not observe the saturation because of the kink instability, 
which will be discussed in detail in Section \ref{sec:sec4}. 

\subsection{\label{sec:sec3.3}Temporal Evolution of Radius}

\begin{figure}[t]
 \centering
  \includegraphics[width=8.cm,clip]{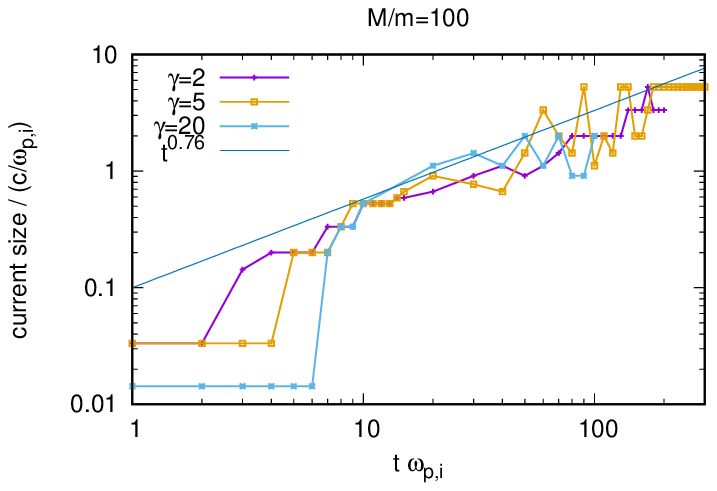}
  \includegraphics[width=8.cm,clip]{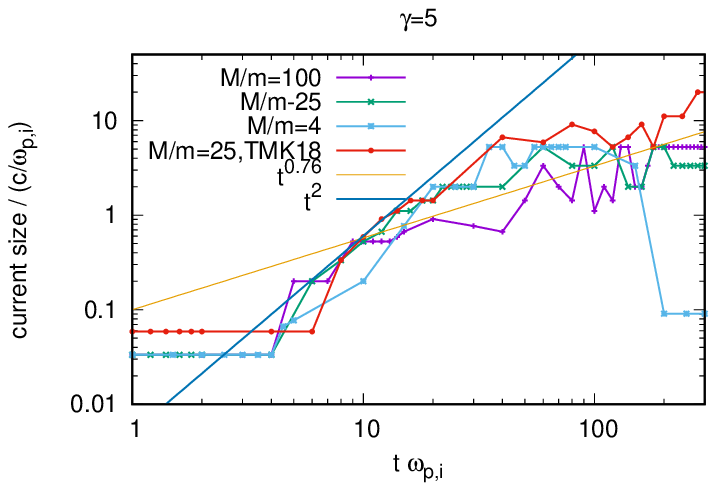}
  \caption{Temporal evolution of filaments' radius. 
           Left: Lorentz factor dependence in the case of $M/m=100$ (runA,B,C). 
           Right: Mass ratio dependence in the case of $\gamma = 5$ (runB,B1,B2,Ba).  
           }
  \label{fig:3.3.1}
\end{figure}

The left panel of Figure \ref{fig:3.3.1} is the temporal evolution of the ion filament radii 
with their initial Lorentz factor varying in the case of $M/m=100$ (runs A,B,C). 
In this paper, 
the filament radius is determined as half of the peak wavelength of the 2D energy spectrum of the ion velocity 
in a plane perpendicular to the initial beam direction. 
This indicates that 
the evolution of the radius is nearly independent of the initial Lorentz factor 
though the timestep of the initial rapid increase of their radius becomes slower as the Lorentz factor increase 
because of the longer growth time. 
The temporal evolution of the radius $R$ in the coalescence phase can be fitted by: $R \propto t^{0.76}$. 

The right panel of Figure \ref{fig:3.3.1} is the temporal evolution of the ion filament radii 
with varying their mass ratio $M/m$ in the case of $\gamma=5$ (runs B,B1,B2,Ba). 
It shows that 
the temporal evolution of the filament radius can be in general described by a power law, $R \propto t^{n}$. 
The temporal evolution of the radius becomes shallower in the later phase around $t > 100 \omega_{p,i}$. 
This is because each filament covers more than 10 \% of the numerical box in this phase 
and the effects of the periodic boundary cannot be neglected. 
In particular, 
it indicates that 
the radius curve becomes steeper when the mass ratio is small, $M/m \lesssim 25$, 
and shallower when the mass ratio is large, $M/m = 100$,
which can be written as:
\begin{align}
  R \propto
  \begin{cases}
    t^{2} & (\text{if} \ M/m \lesssim 25),
    \\
    t^{0.76} & (\text{if} \ M/m = 100).
  \end{cases}
  \label{eq:3.3.1}
\end{align}

\subsection{\label{sec:sec3.4}Temporal Evolution of Anisotropy}

\begin{figure}[t]
 \centering
  \includegraphics[width=8.cm,clip]{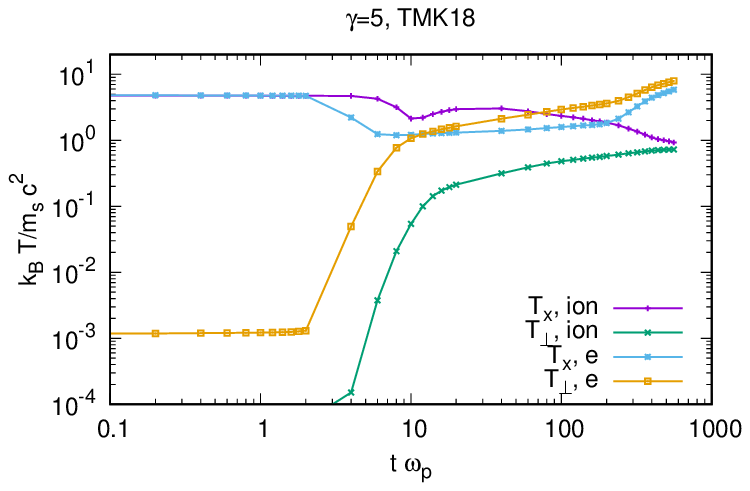}
  \includegraphics[width=8.cm,clip]{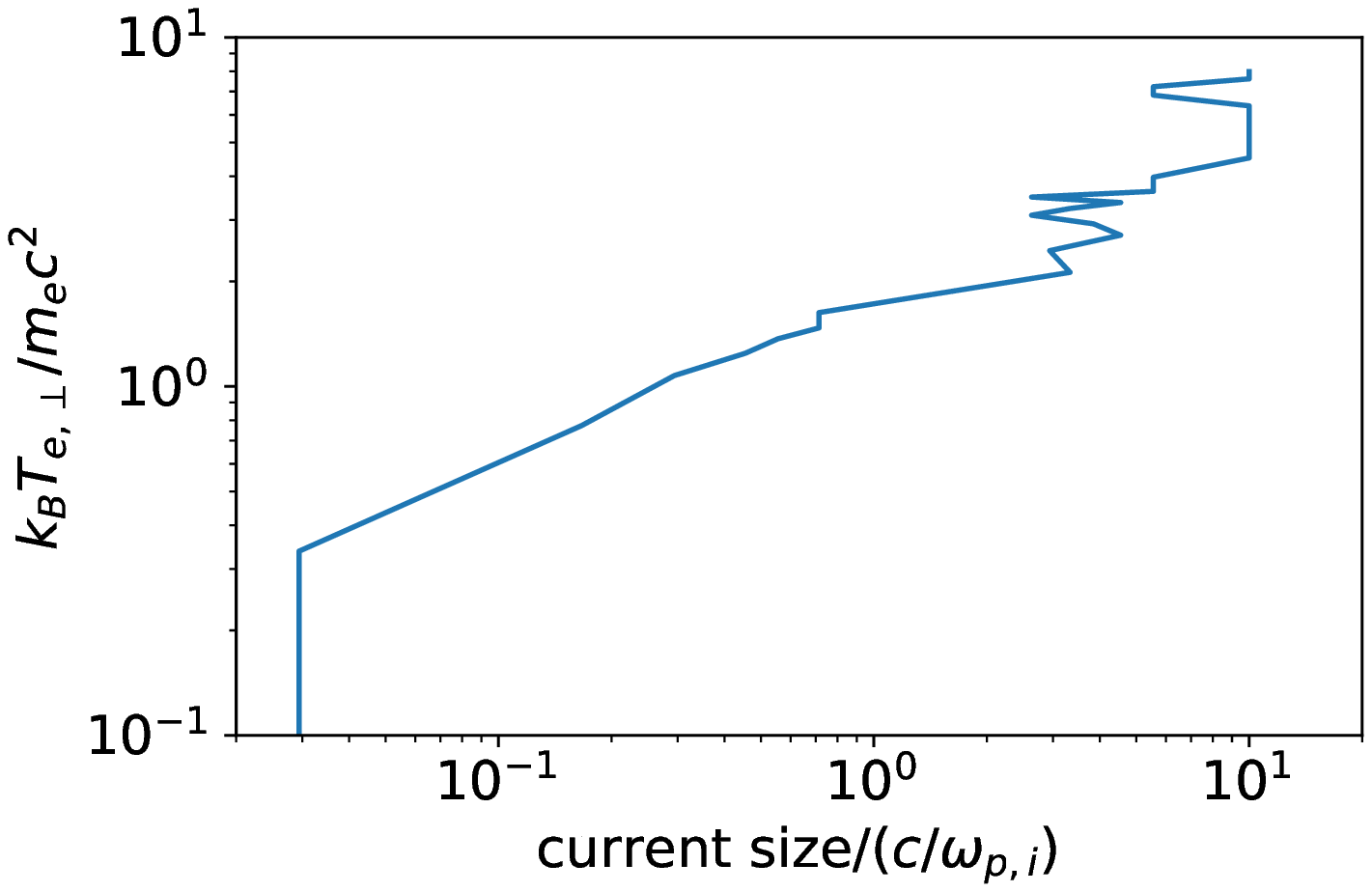}
  \caption{Left: Temporal evolution of temperature of ions and electrons of runBa (TMK18). 
           Right: Plot of electron perpendicular temperature in terms of the ion current filament radius. 
           }
  \label{fig:3.4.1}
\end{figure}

The temporal evolution of temperature of ions and electrons of runBa (TMK18) is presented in Figure \ref{fig:3.4.1}. 
Initially, they only have the parallel temperature from counter-beam distribution, 
and a very small perpendicular temperature from numerical heating
\footnote{
The parallel and perpendicular temperatures are 
measured in terms of the initial beam direction (x-direction), 
but the not magnetic field. 
}
. 
The figure shows that 
the parallel temperature of the ions and electrons starts to reduce 
when they reach the particle-limit current. 
This is because the particle-limit phase is equivalent to the filament coalescence phase 
if the initial anisotropy is sufficiently large. 
Interestingly, 
the temperature evolution of ions and electrons after the particle limit shows a different behavior. 
Electrons become nearly isotropic immediately after reaching the particle-limit current, 
and continue to be isotropic. 

the ion perpendicular temperature also starts to grow during electron's particle limit 
because of turbulent magnetic field generated from electron current filaments. 
However, the ions do not become isotropic even after reaching ion particle-limit current. 
We consider that 
this is because the magnetic at this stage field was too weak to bend the ions' trajectory; 
In contrast, 
the electrons were magnetized because of its small mass. 
This allows their temperature to grow nearly isotropically even after ion filament coalescing phase, 
which can also be observed in the right panel of Figure \ref{fig:3.4.1} 
where electrons' perpendicular temperature increasing with ion filament radius in the power law relation
\footnote{
Note that the electron temperature is not perfectly isotropic but perpendicular temperature is a little larger. 
We consider that 
this is due to the nature of the current filament coalescing processes  
that occur only with current filaments flowing in the same direction. 
This allows strong heating in the perpendicular direction to the current flow. 
}
. 
The fluctuations in beam direction (x-direction) play an important role in the electrons' perpendicular 
temperature evolution. 
In TMK18, it was reported that 
the electron's perpendicular temperature stayed cold in the case of 2D simulation 
with an out-of-plane beam direction. 
The difference comes from the existence of a fluctuation in the beam direction, 
which scatters electrons
through the electrostatic field resulting from these fluctuations 
and causes effective heating. 
This fact claims that 
we need greater caution when using the 2D approximation of plasma simulations, such as PIC, 
and applying the obtained result to the 3D case. 

\section{\label{sec:sec4}Relativistic Magnetohydrodynamic (MHD) Kink Instability}

In this section, 
the kink instability in this simulation is considered. 
We derive the MHD kink instability in Weibel filaments using the energy principle~\citep{1960AnPhy..10..232N,2006ApJ...641..978M}. 
First, 
we discuss the relativistic energy principle following nonrelativistic work by~\citet{2005ppfa.book.....K}. 
Note that our treatment is not covariant but considers only the fluid comoving frame, 
similar to a work by \citet{2018MNRAS.476.4263T}. 
In this section, 
we consider an MHD plasma for simplicity. 
Possible kinetic effects will be discussed in Section \ref{sec:sec6}. 

First, the total energy in a relativistic MHD plasma can be written as: 
\begin{equation}
  {\cal E} = \int d x^3 T_{00} = \int dx^3 \left[\rho h \gamma^2 - p + \frac{B^2 + E^2}{8 \pi} \right]
  .
  \label{eq:4.1}
\end{equation}
Assuming the ideal equation of state: $h = 1 + (p/\rho) [\Gamma/(\Gamma - 1)]$, 
this can be rewritten as: 
\begin{equation}
  {\cal E} = \int dx^3 \left[\rho \gamma + \rho \gamma (\gamma - 1) + \frac{(\gamma^2 - 1) \Gamma + 1}{\Gamma - 1} p + \frac{B^2 + E^2}{8 \pi} \right]
  .
  \label{eq:4.2}
\end{equation}
We consider an energy change from a static state, 
so that the Lorentz factor $\gamma$ can be replaced by $1 + \dot{\xi}^2/2$ 
where $\xi$ is a small displacement.  
The above equation reduces to: 
\begin{equation}
  {\cal E} = \int dx^3 \left[\rho h \dot{\xi}^2 + \rho + \frac{p}{\Gamma - 1} + \frac{B^2 + E^2}{8 \pi} \right] + O(\dot{\xi}^3)
  . 
  \label{eq:4.3}
\end{equation}
This shows that 
we can use the same discussion by \citet{2005ppfa.book.....K} 
formally by just modifying the kinetic energy from $\rho_0 \dot{\xi}^2/2$ to $\rho h \dot{\xi}^2$ 
\footnote{
This expression seems different from the nonrelativistic kinetic energy by a factor of 2. 
This is because the nonrelativistic density is measured in the laboratory frame, 
and is given as $\rho_{\rm NR} \simeq \rho (1 + \dot{\xi}^2/2)$, 
as discussed by~\citet{1959flme.book.....L}. 
In addition, we should subtract the rest mass energy ($\rho \gamma$) from the energy density. 
This results in the nonrelativistic expression. 
}
.
Note that the additional enthalpy $h$ formally includes the effect of the Lorentz factor of the electrons and positively charged particles 
moving with a relativistic velocity. 
Hence, it is formally able to consider the direction of particle counterflows 
if we consider the anisotropic temperatures $T_{||},\ T_{\perp}$ 
that are responsible for the Weibel instability. 
In this section, however, 
we only consider an isotropic temperature for simplicity. 

Following \citet{2006ApJ...641..978M}, 
we consider a cylindrical and symmetric current filament. 
Using the toroidal magnetic field ${\bf B} = B(r) {\bf e}_{\theta}$ and the axial current density ${\bf J} = J(r) {\bf e}_z$, 
the Amp\`ere's law can be written as: 
\begin{equation}
  4 \pi J = \frac{1}{r} \frac{d}{d r} (r B) c
  .
  \label{eq:4.5}
\end{equation}
In the following, 
we assume a homogeneous current $I_0$ in the filament, $r < R$, 
where $R$ is the filament radius. 
From Equation (\ref{eq:4.5}), 
the root-mean-square value of the magnetic field is given as: 
\begin{equation}
  B(r < R) = \frac{\sqrt{2} I_0}{c} \frac{r}{R^2}
  ,
  \label{eq:4.6}
\end{equation}
and the magnetic field outside the current is 
\begin{equation}
  B(r > R) = \frac{\sqrt{2} I_0}{c r}
  , 
  \label{eq:4.7}
\end{equation}
where the reduction of $\sqrt{2}$ is due to the assumption of the presence of a strongly fluctuating magnetic field. 

The static equilibrium is kept by the pressure balance:
\begin{equation}
  \nabla p_{g} = {\bf J} \times {\bf B}
  .
\end{equation}
In the following, 
we consider a small perturbation on this static equilibrium state, 
and the growth of an MHD kink instability. 

In general, 
the second-order potential energy change by a displacement $\xi$ is given as: 
\begin{equation}
  \delta W^{(2)} = \frac{1}{2} \int d x^3 \left[ 
    \frac{Q^2}{4 \pi} + {\bf J} \cdot (\xi \times {\bf Q}) + \Gamma p_{\rm g} (\nabla \cdot \xi)^2 
    + (\xi \cdot \nabla p_{\rm g}) (\nabla \cdot \xi)
  \right]
  ,
  \label{eq:4.8}  
\end{equation}
where ${\bf Q} \equiv \nabla \times (\xi \times {\bf B})$. 
Assuming 
\begin{equation}
  \xi = {\rm Re} \ \left[ (\xi_r, \ i \xi_{\theta}, \ i \xi_z) \ e^{i (m \theta + k z)} \right]
  ,
  \label{eq:4.9}  
\end{equation}
~\citet{1960AnPhy..10..232N} found that 
the minimum $\delta W$ can be obtained if 
\begin{align}
  \xi_r &=  \xi
  \label{eq:4.10}
  ,
  \\
  \xi_{\theta} &=  \frac{i}{m} \left[ \frac{d}{d r}(r \xi) - \frac{k^2 r^4}{k^2 r^2 + m^2} \ \frac{d}{d r} \left( \frac{\xi}{r} \right)  \right]
  \label{eq:4.11}
  ,
  \\
  \xi_z &=  \frac{i k r^3}{k^2 r^2 + m^2} \ \frac{d}{dr} \left( \frac{\xi}{r} \right)
  \label{eq:4.12}
  , 
\end{align}
where $\xi$ is a function of $r$. 
The condition $\nabla \cdot \xi = 0$ is satisfied. 
In this case, the $\delta W$ can be written as: 
\begin{equation}
  \delta W^{(2)} = \frac{\pi}{2} \int^{\infty}_0 g \xi^2 dr
  ,
  \label{eq:4.13}  
\end{equation}
where 
\begin{equation}
  g = \frac{1}{r} \frac{m^2 B^2}{k^2 r^2 + m^2} + \frac{m^2 B^2}{r} - 2 \frac{B}{r} \frac{d}{d r} (r B) + m^2 \frac{d}{d r} \left[ \frac{B^2}{k^2 r^2 + m^2} \right]
  .
  \label{eq:4.14}  
\end{equation}
For the kink instability, 
we consider the case $m = 1$. 
We also assume that the trial displacement vector $\xi$ is a constant, 
which allows us to reduce the complexity of the analysis. 
Although the obtained growth rate is not an exact value of the growth rate, 
obtaining an analytic expression of the growth rate provides a fruitful physical insight into this instability in the relativistic regime. 
The above procedure reduces Equation (\ref{eq:4.14}) to 
\begin{equation}
  g = - \frac{k^2 r}{k^2 r^2 + 1} B_{\theta}^2
  .
  \label{eq:4.15}  
\end{equation}

\begin{figure}[t]
 \centering
  \includegraphics[width=8.cm,clip]{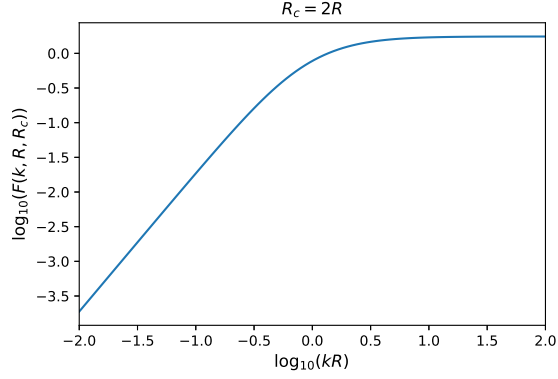}
  \caption{$\delta W^{(2)}$ when $R_{\rm c} = 2 R$. 
           }
  \label{fig:3.1}
\end{figure}

To proceed to the calculation of Equation (\ref{eq:4.14}), 
Equations (\ref{eq:4.6}) and (\ref{eq:4.7}) are substituted into the above equations. 
In this paper, we consider multiple filaments resulting from the Weibel instability 
and there are finite distances between them. 
Therefore, we set a new parameter $R_{\rm c}$ as half the distance between these filaments, 
and we use this as an upper value of integration in Equation (\ref{eq:4.13}). 
We obtain 
\begin{align}
  \delta W^{(2)} &= - \frac{\pi I_0^2 \xi^2}{2 R^2 c^2} F(k,R,R_{\rm c})
  \label{eq:4.16a}
  ,
  \\
  F(k, R, R_{\rm c}) &= 
  \left[ 
    1 - \frac{\ln [1 + k^2 R^2]}{k^2 R^2} + k^2 R^2
    \left(
      \ln \left[\frac{1 + k^2 R^2}{k^2 R^2} \right]
    - \ln \left[\frac{1 + k^2 R_{\rm c}^2}{k^2 R_{\rm c}^2} \right]
    \right)
  \right]
  .
  \label{eq:4.16b}  
\end{align}
In the limit of a small and large value of $k$, 
the function $F(k, R, R_{\rm c})$ becomes 
\begin{align}
  F(k, R, R_{\rm c}) \rightarrow 
  \begin{cases}
    \left[1/2 + 2 \ln (R_{\rm c}/R) \right] k^2 & (\text{if} \ k R \ll 1)
    \\
    2 - R^2/R_{\rm c}^2 &  (\text{if} \ k R \gg 1)
  \end{cases}
  \label{eq:4.17}
  ,
\end{align}
and the detailed dependence on $k$ is given in Figure \ref{fig:3.1}. 
Equations (\ref{eq:4.16a}) and (\ref{eq:4.16b}) show that 
$\delta W^{(2)}$ is always negative, 
which means that these filaments are unstable for the MHD kink instability for any wavelength. 

Next, we calculate the kinetic energy. 
According to \citet{2005ppfa.book.....K}, 
we can estimate the upper limit of the growth rate $\Gamma_{\rm grow}$ of perturbations as the following expression:
\begin{equation}
  \Gamma_{\rm grow} \lesssim \sqrt{\frac{\delta W^{(2)}}{K}}
  ,
  \label{eq:4.18}  
\end{equation}
where $K$ is the kinetic energy given as: 
\begin{equation}
  K \equiv \int \rho h (\xi_r^2 + \xi_{\theta}^2 + \xi_{\phi}^2)
  .
  \label{eq:4.19}
\end{equation}

Note that the kinetic energy includes the density $\rho$ and the pressure $p_{\rm gas}$ in its enthalpy 
for the following reason. 
The equation of motion of RMHD can be written as: 
\begin{equation}
  \rho h \frac{\partial^2}{\partial t^2} \xi = - \nabla p + {\bf j} \times {\bf B}
  \label{eq:4.4}
  , 
\end{equation}
in a second-order accuracy in $\xi$. 
This means that 
the discussion deriving the growth rate of the nonrelativistic case in \citet{2005ppfa.book.....K} can also be formally applied 
in the relativistic case 
if we replace the kinetic energy from $\rho \dot{\xi}^2/2$ to $\rho h \dot{\xi}^2$. 

\begin{figure}[t]
 \centering
  \includegraphics[width=7.cm,clip]{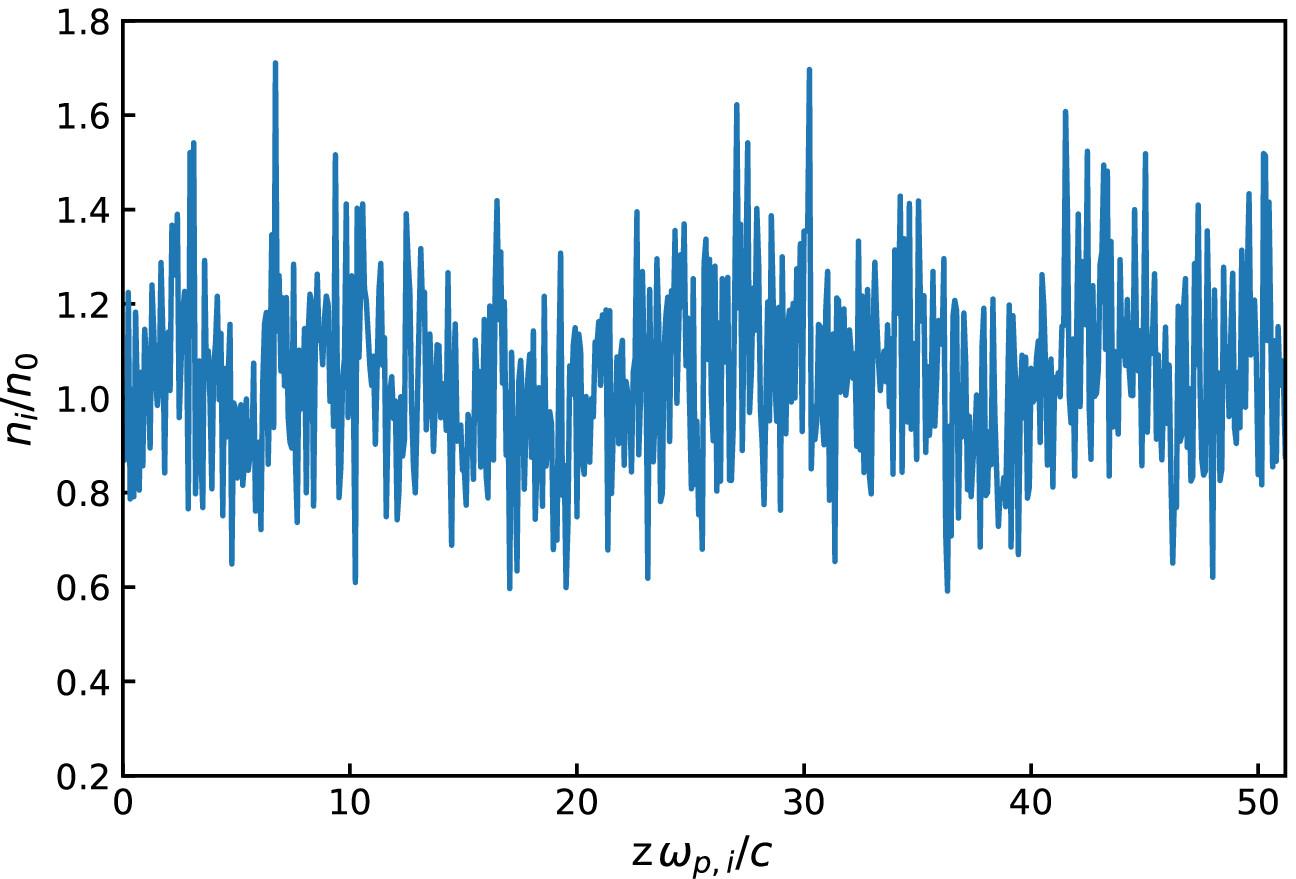}
  \includegraphics[width=8.cm,clip]{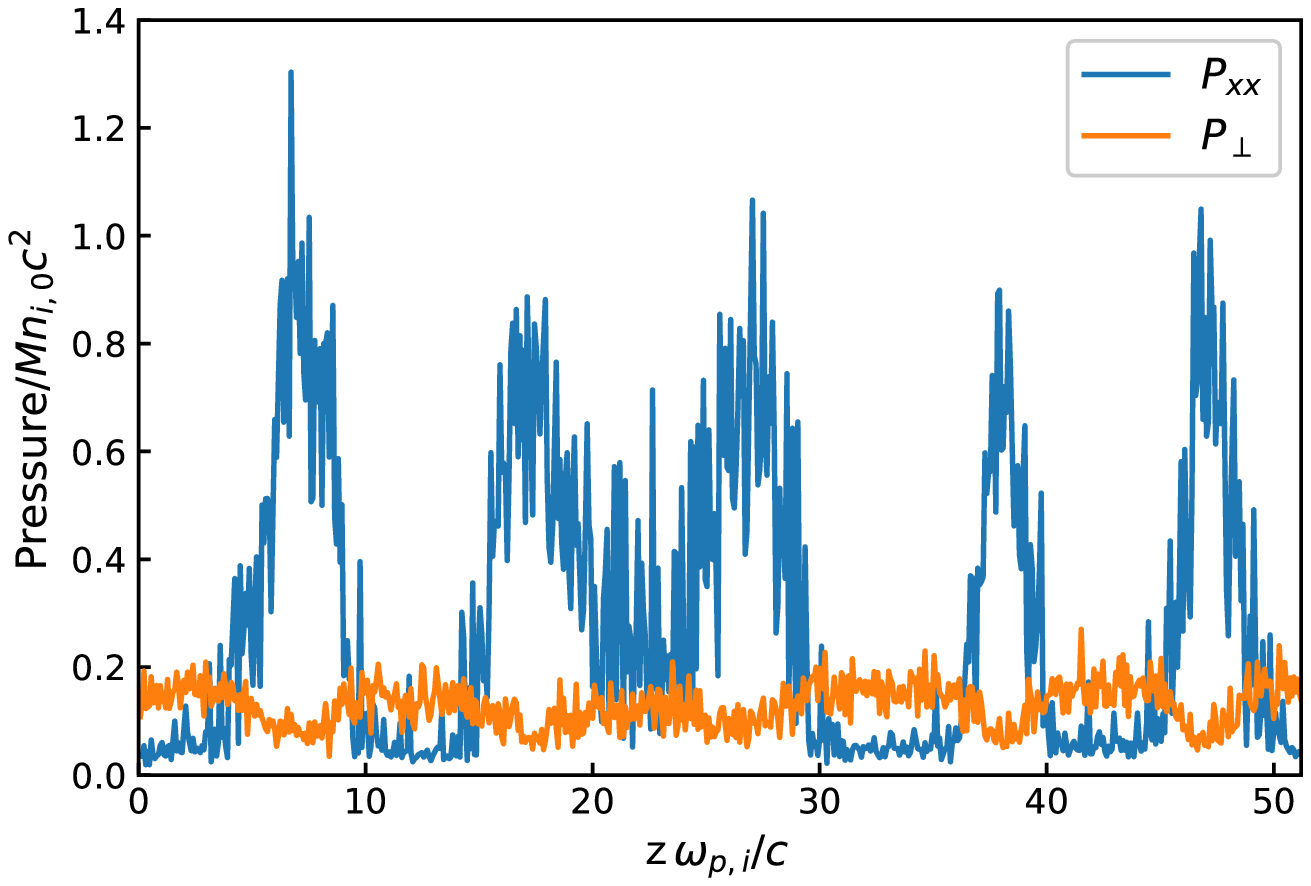}
  \includegraphics[width=10.cm,clip]{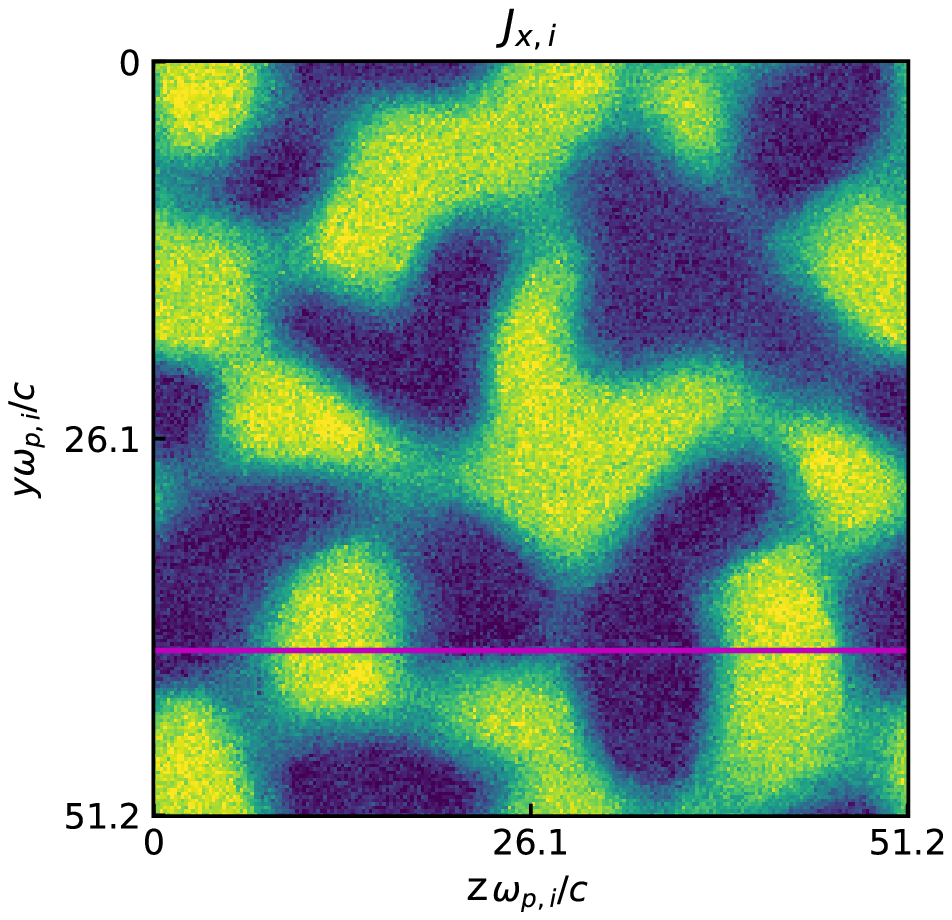}
  \caption{Top Left: One-dimensional (1D) ion density profile, 
           Top Right: 1D ion pressure profile. 
           They are measured at $t=200 \omega_{p,i}^{-1}$ and $x=2.52 c/\omega_{p,i}$, 
           and the line is taken from the magenta line in the bottom panel 
           which is the 2D ion current density profile at $t=200 \omega_{p,i}^{-1}$ and $x=2.52 c/\omega_{p,i}$. 
           In the panel, the yellow and blue regions show positive and negative current densities, respectively. 
           }
  \label{fig:3.2.lin}
\end{figure}

The top panel of Figure \ref{fig:3.2.lin} shows the profiles of the ion density and pressure along a magenta line listed in the bottom panel of Figure \ref{fig:3.2.lin}. 
They show that the density is almost uniform in all regions, 
and pressure is constant with different values according to whether it is inside or outside of the filaments
\footnote{
The pressure in the bottom panel of Figure \ref{fig:3.2.lin} is measured in the ion rest frame. 
Inside the filaments, 
the ions maintain the initial beam velocity due to the collisionless property 
and the temperature in the x-direction becomes cold. 
In contrast, 
in the contact region, 
there are two ion components with opposite flow directions, 
which results in a very high temperature in the x-direction. 
}, 
which we set as $\rho_0, p_{\rm in}, p_{\rm out}$, respectively. 
Substituting Equations (\ref{eq:4.10}), (\ref{eq:4.11}), and (\ref{eq:4.12}), 
the kinetic energy becomes 
\begin{equation}
  \frac{\pi \xi^2}{2 k^2} \left[ \rho_0 h_{\rm out} \{5 k^2 R_{\rm c}^2 - 3 \ln (k^2 R_{\rm c}^2 + 1)\} 
    - 4 \Delta p \{5 k^2 R^2 - 3 \ln (k^2 R^2 + 1)\} \right]
  ,
  \label{eq:4.20}
\end{equation}
where $h_{\rm out}$ is the enthalpy in the contact region of the filaments, 
and $\Delta p \equiv p_{\rm out} - p_{\rm in}$. 
From Equations (\ref{eq:4.16a}), (\ref{eq:4.16b}), and (\ref{eq:4.20}), 
the growth rate is given as:
\begin{equation}
  \Gamma_{\rm grow} = \left[ \frac{I_0^2 k^2}{R^2 c^2} \frac{1 - \ln[1 + k^2 R^2]/k^2 R^2] 
    + k^2 R^2 \left\{ \ln[1 + 1/k^2 R^2] - \ln [1 + 1/k^2 R_{\rm c}^2] \right\} }
    {\rho_0 h_{\rm out} \{5 k^2 R_{\rm c}^2 - 3 \ln (k^2 R_{\rm c}^2 + 1)\} 
    - 4 \Delta p \{5 k^2 R^2 - 3 \ln (k^2 R^2 + 1)\}} \right]^{1/2}
  \label{eq:4.21}
\end{equation}
In the long and short wavelength limit, 
this reduces to
\begin{align}
  \Gamma_{\rm grow} \rightarrow 
  \begin{cases}
    \frac{c_{\rm A} k}{2 \sqrt{2}} \left(\frac{R}{ R_{\rm c}} \right) \sqrt{\frac{1 + 4 \ln(R/R_{\rm c})}{1 + \frac{4 \Delta p}{\rho_0 h_0} \frac{R_{\rm c}^2 - R^2}{R_{\rm c}^2}}} & (\text{if} \ k R \ll 1),
    \\
    \frac{c_{\rm A}}{\sqrt{10} R_{\rm c}} \sqrt{\frac{2 - (R/R_{\rm c})^2}{1 + \frac{4 \Delta p}{\rho_0 h_0} \frac{R_{\rm c}^2 - R^2}{R_{\rm c}^2}}} &  (\text{if} \ k R \gg 1).
  \end{cases}
  \label{eq:4.22}
\end{align}
\begin{figure}[t]
 \centering
  \includegraphics[width=10.cm,clip]{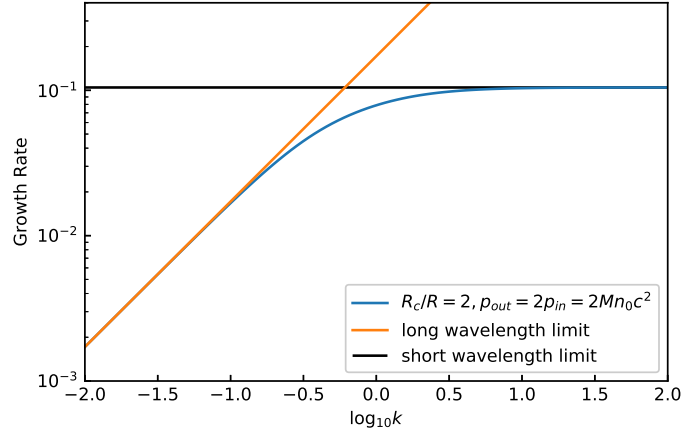}
  \caption{Plot of growth rate in Equation (\ref{eq:4.21}) in terms of wave number $k$ 
           when $R_{\rm c} = 2 R$ and $p_{\rm out} = 2 p_{\rm in} = 2 M n c^2$. 
           }
  \label{fig:3.3}
\end{figure}
Figure \ref{fig:3.3} is a plot of Equation (\ref{eq:4.21}) when  $R_{\rm c} = 2 R$ and $p_{\rm out} = 2 p_{\rm in} = 2 M n c^2$. 
It shows that 
the growth rate reaches its maximum value around $k \simeq c/\omega_{p,i}$, 
the value of which can be well reproduced by the short wavelength limit in Equation (\ref{eq:4.22}). 

From our simulation results, 
we set $R_{\rm c} \simeq 2 R$, $p_{\rm out} \simeq 2 p_{\rm in} \sim 2 M n c^2$, 
the growth rate in the range of wavelength shorter than $R$ becomes 
\begin{equation}
  \Gamma_{\rm grow} \simeq 0.15 \frac{c_{\rm A}}{R}
  ,
  \label{eq:4.23}
\end{equation}
and this shows that 
the growth of the MHD kink instability in current filaments with Alfv\'en current is around 10 times the Alfv\'en crossing time of filaments
\footnote{
Our assumption of a comoving boundary with the current filament is in fact invalid 
and it has a steep velocity shear, 
which would cause another instability, such as the Kelvin-Helmholtz instability. 
However, this was neglected here to avoid a too-complex analysis. 
In addition, 
the radius $R$ is the shortest limit of validity of MHD approximation. 
Hence, 
Equation (\ref{eq:4.23}) also corresponds to the maximum growth rate 
which is valid in the case of the MHD approximation. 
}
. 
Note that 
our expression differs from Equation (14) of \citet{2006ApJ...641..978M} 
because of the modification in Equation (\ref{eq:4.4}). 

Finally, 
this analysis was performed in the fluid rest frame; 
therefore, the growth rate in the laboratory frame could be obtained as: 
\begin{equation}
  \Gamma_{\rm grow, lab} \simeq 0.15 \frac{c_{\rm A}}{\gamma_{\rm fluid} R} 
  ,
  \label{eq:4.24}
\end{equation}
where $\gamma_{\rm fluid}$ is the Lorentz factor of the background fluid. 
This is also similar to the expression obtained in the case of force-free approximation \citep{1999MNRAS.308.1006L} 
although its Lorentz factor is not exactly equal to the fluid one but to the phase velocity of the perturbation. 

\section{\label{sec:sec6}Saturation Mechanisms  -- Suppression of MHD Kink Instability}

\begin{figure}[t]
 \centering
  \includegraphics[width=8.cm,clip]{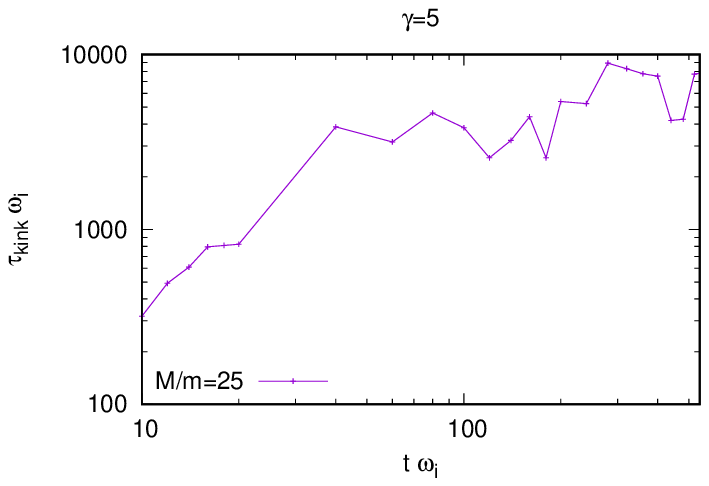}
  \includegraphics[width=8.cm,clip]{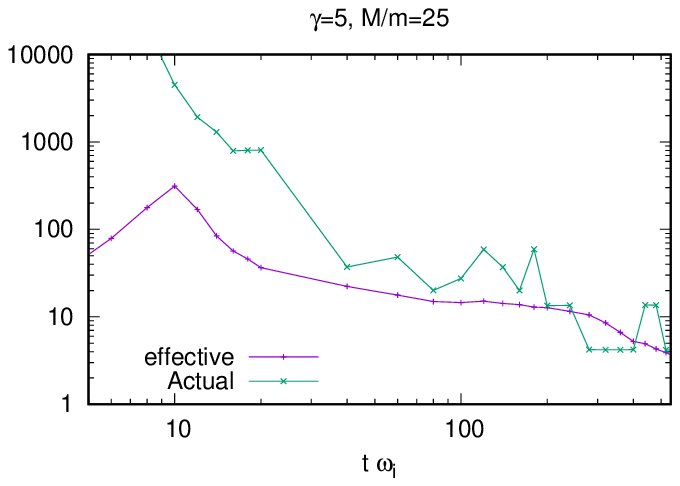}
  \caption{Left: Temporal evolution of the growth time of the MHD kink instability 
           in the case of runBa. 
           Right: Temporal evolution of the effective and measured topological constants of runBa.
           }
  \label{fig:6.2}
\end{figure}

Our simulations indicate that 
the Alfv\'en current limit and isotropization of particles are responsible for the saturation of relativistic Weibel instability.
However, 
it has long been believed that 
the Weibel filaments would suffer from kink instability, 
resulting in saturation of the filament evolution. 
In Section \ref{sec:sec4}, 
we calculated the growth rate of kink instability in the case of Weibel filaments using the energy principle 
given by Equation (\ref{eq:4.24}). 
Left panel of Figure \ref{fig:6.2} plots the growing time of MHD kink instability, the inverse of Equation (\ref{eq:4.24}), in the cases of runBa. 
Note that 
the Alfv\'en velocity was calculated considering the effect of the relativistic pressure, 
which reduces the Alfv\'en velocity more than expected from $\epsilon_{\rm B}$. 
This indicated that 
the MHD kink instability cannot grow in both cases 
because the growth time is much longer than the simulation timescale. 



In addition, 
we consider that 
the formation of a nontrivial 2D structure of current filaments in the plane perpendicular to the beam direction is also important. 
The bottom panel in Figure \ref{fig:3.2.lin} shows that 
the Weibel filaments do not exist as isolated cylindrical filaments 
which was assumed for the analysis of kink instability, 
but as a network of multifilaments. 
This nontrivial network structure would cancel the nonlinear evolution of kink instability, 
and keep the filaments stable. 
To characterize the topology of the current filament network quantitatively, 
we define the following topological variable: 
\begin{equation}
  N_{\rm TPL} \equiv \frac{L_{\rm BD}^2}{4 \pi S_{+/-}}
  ,
  \label{eq:6.1}
\end{equation}
where $L_{\rm BD}$ is the length of the boundary between the current filaments, 
and $S_{+/-}$ is the area of current filaments with a positive/negative current. 
The above new variable reduces to the number of filaments  
if the filaments are isolated cylindrical currents 
with a similar radius
: $L_{\rm BD} \rightarrow \Sigma 2 \pi R = 2 \pi N R$ and $S \rightarrow \Sigma \pi R^2 = \pi N R^2$.
This means that 
$N_{\rm TPL}$ decreases as filaments merge and a network of the filaments is formed. 
The right panel of Figure \ref{fig:6.2} is the temporal evolution of $N_{\rm TPL}$ and the number of pseudocircles 
measured through a typical length scale the value of which is shown in the right panel of Figure \ref{fig:3.3.1}, 
which counts the number of filaments if the filaments are isolated
\footnote{Here, 
$S_{+/-}$ were measured by counting the area 
where the current density is larger than 2.5\% of the averaged current density. 
$L_{\rm BD}^2$ were measured by counting the area with a current density of less than 2.5\% of the averaged current density, 
and divided by $\pi R^2$, where R is the value plotted in Figure \ref{fig:3.3.1}. 
The threshold 2.5\% is comparable to the level of numerical fluctuations. }. 
The figure shows that 
the $N_{\rm TPL}$ is nearly always less than the pseudocircles, 
indicating that the filaments form a network 
and the nonlinear evolution of kink instability cannot grow due to the cancellation in the network structure
\footnote{
Although the effective $N_{\rm TPL}$ becomes partly less than the actual value, 
we consider that this is due to the numerical error for calculating $N_{\rm TPL}$ from our numerical simulation results. 
}
.

Although the above discussion assumes an MHD approximation for the kink instability, 
it is still possible that some kinetic effects play important roles 
because the radius of cylindrical Alfv\'en currents is comparable to the Larmor radius of the generated magnetic field. 
\citet{2018PhRvL.120x5002R} indicated that 
the resonance between the particles' gyromotion and the current filament motion can induce a kinetic kink-type instability. 
The unstable wavelength is predicted as comparable with one gyrorotation length: 
$2 \pi v_{\perp,i}/\omega_B = 2 \pi \sqrt{\gamma/2 \epsilon_B (\gamma - 1)}(v_{\perp,i}/c)(c/\tilde{\omega}_{p,i}) 
= 2 \pi \sqrt{\gamma^2/2 \epsilon_B (\gamma - 1)}(v_{\perp,i}/c)(c/\omega_{p,i})$ 
where $\omega_B = e B / \gamma M c$ is the Larmor frequency and a cold plasma is assumed. 
The growing time is predicted as: $\sqrt{\gamma} c/ (\kappa v_{\perp, i} \omega_{p,i}) = c/ (\kappa v_{\perp, i} \tilde{\omega}_{p,i})$, 
where $\kappa$ is the screening factor of current by hot electrons. 
The relativistic modification from the Lorentz dilation was applied in the above expressions. 
In our simulation (runBa), 
the growing time reduced to around $11 \omega_{p,i}^{-1}$ and the unstable wavelength around $90 c/\omega_{p,i}$, 
where we used $\gamma = 5$, $\epsilon_B = 0.015$, $\kappa = 0.2$, and $v_{\perp,i} = c$. 
This means that 
it might be able to grow, but our numerical box is not large enough to check it, 
although it might be suppressed by a similar effect from the network structure discussed above. 
Unfortunately, 
the present numerical resource does not allow us to check the above mechanism with a kinetically sufficient resolution ($\Delta \lesssim 0.1 c/\omega_{p,e}$), 
and we leave this as our future work
\footnote{
This topic has now been actively studied, 
and the recent developments can be found in, e.g., \citet{2018PhRvL.121x5101A} and \citet{2019Galax...7...29N} 
which discuss the importance of the kinetic Kelvin-Helmholtz instability, the application to AGN jets, and so on. 
}
. 

\section{Summary}

In this paper, 
a detailed explanation of our previous work TMK18 was provided. 
It was found that 
the magnetization parameter $\epsilon_{\rm B}$ was nearly insensitive to the initial beam Lorentz factor and mass ratio, 
saturating around 0.01 to 0.03. 
It was also found that 
the evolution of the relativistic Weibel instability can be divided into six phases 
which are closely related with the beam plasma current limit: particle and Alfv\'en limit currents. 
The important finding was that 
the relativistic MHD kink instability was analyzed using the energy principle method. 
The obtained growth time is much longer than our simulation time. 
This indicates that 
the relativistic MHD kink instability may not suppress the relativistic Weibel instability, 
and the resulting magnetic field can be 
a seed of a large scale MHD instability. 

\acknowledgments

We would like to thank Takanobu Amano, Masahiro Hoshino 
for many fruitful comments and discussions. 
This research used computational resources of the K computer provided by the RIKEN Advanced Institute for Computational Science through 
the HPCI System Research project (Project ID:hp160121, hp170125, hp170231). 
This work is supported in part by the Postdoctoral Fellowships by the Japan Society for the Promotion of Science No. 201506571 (M. T.), and JSPS KAKENHI grant No. 17H02877 (Y. M.).

%
%



\end{document}